 \pdfoutput=1

\documentclass[journal]{IEEEtran}
\usepackage{nomencl}

\usepackage{cite}

\usepackage{multirow}
\usepackage{lineno}

\ifCLASSINFOpdf
   \usepackage[pdftex]{graphicx}
\else
\fi
\usepackage{amsmath,amsfonts,amssymb}
\usepackage{mathtools}

\usepackage{array}
\usepackage{amsmath}

\usepackage[fleqn]{nccmath}
\usepackage{amsfonts}

\usepackage{array}
\usepackage{booktabs}
\usepackage{makecell}
\usepackage[dvipsnames]{xcolor}
\usepackage[hyphens,spaces]{url}
\usepackage[ruled,vlined]{algorithm2e}
\usepackage[flushleft]{threeparttable}

\begin{document}
%
\title{Distributionally Robust Joint Chance-Constrained Optimization for Networked Microgrids Considering Contingencies and Renewable Uncertainty}

\author{Yifu~Ding,~\IEEEmembership{Student Member,~IEEE,}
        Thomas~Morstyn,~\IEEEmembership{Member,~IEEE,}\\
        Malcolm~D. McCulloch,~\IEEEmembership{Senior~Member,~IEEE}
\thanks{This work is supported in part by the Engineering and Physical Sciences Research Council under Grant (EP/R030111/1) Robust Extra Low Cost Nano-grids (RELCON).}
\thanks{Y. Ding and M. McCulloch are with the Department
of Engineering Science at the University of Oxford, Oxford OX1 2UD, United Kingdom.
 (e-mail: yifu.ding@eng.ox.ac.uk, malcolm.mcculloch@eng.ox.ac.uk).}
\thanks{T. Morstyn is with the School of Engineering Science at the University of Edinburgh, Edinburgh EH9 3JW, United Kingdom. (e-mail: Thomas.Morstyn@ed.ac.uk)}}

\maketitle


\begin{abstract}
In light of a reliable and resilient power system under extreme weather and natural disasters, networked microgrids integrating local renewable resources have been adopted extensively to supply demands when the main utility experiences blackouts. However, the stochastic nature of renewables and unpredictable contingencies are difficult to address with the deterministic energy management framework. The paper proposes a comprehensive distributionally robust joint chance-constrained (DR-JCC) framework that incorporates microgrid island, power flow, distributed batteries and voltage control constraints. All chance constraints are solved \textit{jointly} and each one is assigned to an \textit{optimized} violation rate. To highlight, the JCC problem with the \textit{optimized} violation rates has been recognized as NP-hard and challenging to solve. This paper proposes a novel evolutionary algorithm that successfully solves this problem and reduces the solution conservativeness (i.e. operation cost) by around 50\% compared with the baseline \textit{Bonferroni Approximation}. We construct three data-driven ambiguity sets to model uncertain solar forecast error distributions. The solution is thus robust for any distribution in sets with the shared moments and shape assumptions. The proposed method is validated by robustness tests based on these sets and firmly secures the solution robustness.

\end{abstract}

\IEEEpeerreviewmaketitle

\section{Nomenclature}
\begin{table}[!h]
\label{table_example}
\normalsize
\renewcommand\arraystretch{0.95}
\begin{tabular}{lp{2.5in}}
\textit{A.}& \textit{Set and index}\\
$\mathcal {T}, t$& Set, index of timesteps\\
$\mathcal{B}, b$& Set, index of buses\\
$\mathcal {S}, s$& Set, index of distributed assets (i.e. storages, PV panels, loads)\\
&\\
\textit{B.}& \textit{Parameters and variables}\\
$m_{u}$ & Cost factor of grid power\\
$m_{s}$ & Cost factor of solar generation curtailment\\
$m_{l}$ & Cost factor of load curtailment\\
$m_{r}$ &Cost factor of droop control provision \\
$m_{d}$ &Battery degradation costs \\
$N_{B}$ & Number of network buses\\
\end{tabular}
\end{table}

\begin{table}[!h]
\label{table_example}
\normalsize
\renewcommand\arraystretch{0.95}
\begin{tabular}{lp{2.5in}}
$N_{L}$ & Number of lines \\
$N_{S}$ & Number of distributed batteries\\
$N_{D}$ & Number of loads\\
$N_{PV}$ & Number of solar PV panels\\
$N_{it}$ &Maximum number of evolution iterations\\
$N_{c}$ &Number of single chance constraints\\
$N_{s}$ &Number of forecast error samples\\
$N_{p}$ &Number of individuals in one generation\\
$\eta_{dis}, \eta_{ch}$&Battery discharging / charging efficiency\\
$\overline {v}, \underline {v}$ & Maximum / minimum bus voltage\\
$v$ & Voltage magnitude of buses\\
$\overline{SoC}, \underline{SoC}$&Maximum / minimum state of charge (SoC)\\
$\overline{P^c}, \overline{P^d}$&Maximum discharging / charging power\\
$P^{d}, P^{c}$ & Battery discharging / charging power \\
$P^{l}$& Demands at each bus\\
$P^{l'}$& Supplied loads at each bus\\
$P^{cl}$&Critical loads at each bus\\
$\overline{E^{s}}$& Battery energy capacity\\
$\mu_{pv}$ &Mean vector of solar forecast errors\\
$\Sigma_{pv}$& Covariance matrix of solar forecast errors\\
$\overline{\epsilon}, \underline{\epsilon}$&Upper / lower bounds of violation rates of chance constraints\\
$\epsilon_i$& Violation rate of single chance constraints\\
$\epsilon_j$& Violation rate of joint chance constraints\\
$d^{s}$& Droop provision coefficients\\
$P^{pv}$ & Forecast solar power at each bus\\
$P^{pv'}$ & Consumed solar power at each bus\\
$\tilde{P^{pv}}$&Solar power forecast errors\\
$P^{inj}$ &Net injected power at each bus\\
$P^{g}$ & Imported grid power\\
$R^{up}$ & Upward droop provision from batteries\\
$E^{s}$& SoC of distributed batteries\\
$r$ & Ratio to measure the solution convergence\\ 
$r_{thr}$ & Threshold ratio for the termination condition\\ 
$\textbf{V}_{o}, \textbf{I}_{o}$ & Voltage / current operation point vector\\ 
$\textbf{Y}$ & Network admittance\\ 
$\textbf{C}$ & Sparse matrices to map the distributed assets\\ 
$\textbf{P}$ & Individuals (population) in one generation
\end{tabular}
\end{table}

\section{Introduction}

\IEEEPARstart{T}he reliability of the power system under the impact of increasing renewable penetration and extreme weather conditions is a rising concern. In the developed world, the decreasing number of dispatchable fossil-fuel power plants and reduced system inertia render the power system more vulnerable to natural disasters\cite{Johnson2019}. Recent examples include the rolling blackouts across California due to the wildfire\cite{Penn2020} and disastrous power outages in Texas due to the extremely cold weather \cite{Levin2022}. In the developing world, increasing electricity demand and aging infrastructure result in frequent power outages. In sub-Saharan African countries, the outage time of public utility is commonly around 10\%, and even reaches 50\% in some instances \cite{Lee2018}.

To tackle utility failures and power cuts, microgrids aggregate local renewable energy resources and loads in a small network, and operate flexibly with or without the grid connection, thanks to modern inverter-based design \cite{Pogaku2007}. Microgrids can thus supply loads when the grid experiences scheduled under-frequency load shedding (UFLS)\cite{Teng2016} or unpredictable power cuts. This islanding capability is incorporated into the power scheduling of microgrids \cite{Khodaei2014, Liu2017}. Smart load shedding is also employed to mitigate power imbalances in microgrids. For example, refs. \cite{Li2011a, Bahrami2018} model users' utility functions using different appliances so that flexible loads can be shifted efficiently for peak shaving in distributed power networks. 

\textcolor{black}{On the other hand, intermittent renewable resources such as solar power pose challenges to short-term power system operations. The imperfect forecast brings uncertainty, which could result in network constraint violations and high power losses \cite{Fortenbacher2017d}. Stochastic and robust optimization have been proposed to address the uncertainty. While robust optimization based on the worst-case scenario leads to an overly-conservative and cost-prohibitive solution, the chance-constrained (CC) formulation, as one of the predominant stochastic approaches, can directly control the system reliability to a predefined level and decide the optimal cost. The CC formulation of the optimal power flow (OPF) was first proposed in \cite{Zhang2011}, incorporating a series of single network chance constraints (CCs) pertaining to voltage and power limits. }

\textcolor{black}{The most intuitive way to solve the CC problem is the scenario-based approach. As the exact solution of a CC problem is unattainable, this approach solves a great number of problem scenarios randomly drawn from the uncertainty distribution. To secure the estimation confidence level $1-\beta$, the number of random samples should be at least $\frac{2}{\epsilon}(\ln\frac{1}{\beta}+n)$, given the violation rate $\epsilon$ and the dimension of decision variables $n$ \cite{Calafiore2006b}.}

A more effective alternative is distributionally robust optimization (DRO). This approach constructs a set based on historical data - termed the \textit{ambiguity set} - including all possible uncertainty distributions. The formulation thus ensures constraints are satisfied for any distribution in the ambiguity set built upon distribution moments and shape information. The problem can be solved by being recast into tractable formulations, including linear programming (LP), semidefinite programming (SDP) and second-order conic programming (SOCP) depending on the degree of approximation \cite{Delage2010}. However, defining an ambiguity set to characterize uncertain distributions is non-trivial, as one needs to decide the trade-off between solution robustness and conservativeness, while considering the mathematical tractability \cite{Rahimian2019}. Compared with early works using the first two moments (i.e. mean and variance) such as \cite{Zhang2017}, recent works utilize the high-order moments (e.g. skewness\cite{Zhao2020}), structural properties (e.g. unimodal \cite{Pourahmadi2021} and symmetric \cite{Fang2019}) to set tighter bounds. Another kind of method is the moment-free method. Ref. \cite{Guo2019b} constructs a ball space where possible distributions are centred at the reference distribution based on the training samples, and the ball radius is defined by Wasserstein-based distance metrics.  However, such an approach is highly data-intensive and its performance is substantially influenced by the volume of data available \cite{Duan2018}. 

All aforementioned literature \cite{Zhang2017, Qiu2019, Zhao2020, Shi2019, Fang2019, Duan2018, Guo2019b} adopt the single CC formulation, in which each constraint is considered as an independent event with the pre-defined violation rate. However, in most power system applications, the JCC formulation is desired, which means that all constraints should be satisfied simultaneously and use one whole-system reliability metric. For example, a distribution feeder is considered to be reliable \textit{if and only if} all the constraints such as bus voltage limits, power balance are met simultaneously. 

\textcolor{black}{However, solving the JCC problem is notoriously difficult since its DRO reformulation generally results in intractable problems \cite{Hanasusanto2017}. Only a few papers attempted to solve the JCC problems using either the scenario-based or approximation-based methods. The scenario-based method, following the aforementioned principles, solves possible scenarios from historical samples for the JCC problem. Ref. \cite{Vrakopoulou2013} first proposed the JCC-OPF formulation for the transmission network with high wind power penetration. A droop-type function, termed the distribution vector, was introduced to control generators concerning wind power forecast errors \cite{Vrakopoulou2013, Vrakopoulou2012}. These problems were solved in a great number of wind power forecast scenarios. Similar works include \cite{Warrington2012, Roald2018, Vrakopoulou2019} for power networks integrating flexible loads or thermal storages. In general, the scenario-based approach offers a fairly accurate solution given a large volume of samples, but its scalability heavily relies on statistical techniques, such as sample average estimation, to ease the computation burden \cite{Pena-Ordieres2021a}. }

The approximation-based method is decomposing an intractable JCC problem into a series of tractable SCCs, then approximating individual violation rates. The simplest approximation method, termed the \textit{Bonferroni Approximation}, assumes that all individual violation rates are the same and equal to the joint violation rate divided by the number of individual constraints, proposed first in \cite{Nemirovski2006}. This approximation has an extremely conservative assumption that neglects the intersections of constraints and treats all constraints equally. In this case, the solution conservativeness increases with the number of individual chance constraints\cite{Zymler2013}. To reduce the solution conservativeness, ref. \cite{Baker2019} identifies all intersections of constraints using machine learning classification and obtained around 5\% result improvement, compared to the \textit{Bonferroni Approximation} as the baseline. Refs. \cite{Chen2010, Sun2014a} approximate the JCC to conditional value-at-risk (CVaR) constraints and introduce a scaling factor to control the tightness of the approximation. The improvement benchmarked against the \textit{Bonferroni Approximation} is around 8-12\%. However, none of the previous papers try to allocate the optimized violation rates for each constraint. Ref. \cite{Baker2019} concludes these optimal violation rates are challenging to find. Furthermore, ref. \cite{Xie2019} theoretically proves that a JCC problem with the optimized individual violation rates, termed the \textit{optimized Bonferroni Approximation}, is a strongly NP-hard problem.

This paper makes the following contributions which together address the aforementioned challenges:

1) \textcolor{black}{A novel evolutionary algorithm is proposed to solve the JCC problem with optimized individual violation rates, which is an NP-hard problem and challenging to solve. Our method shows around a 50\% reduction in the solution conservativeness (i.e., operation cost) benchmarked against the \textit{Bonferroni Approximation}. This performance is the best to date compared to other approximation-based methods for the JCC problem.  Moreover, these optimized violation rates are interpretable, accurately reflect the sensitivities of corresponding constraints to the operation cost. }

2) The proposed JCC algorithm is tested on three data-driven ambiguity sets, namely, symmetrical, unimodal and symmetrical \& unimodal sets. These ambiguity sets are created and constructed using the empirical solar power forecast errors from a machine learning model to capture accurate statistical characteristics of uncertainty distributions in each time interval.

3) This DR-JCC energy management framework for the networked microgrid incorporates CCs pertaining to the power flow, bus voltage, energy storage power and energy limits. We run the power flow simulations under three uncertain distribution assumptions (i.e. ambiguity sets) to observe the no-violation cases. Only the proposed method can schedule the system to closely meet the reliability requirements, while the SCC and the benchmark case give either unreliable or overly-conservative results.  

The rest of the paper is organized as follows. Section III presents \textcolor{black}{the centralized OPF formulation for a networked DC microgrid}. Section IV demonstrates the essential steps to reformulate the model into a DR-JCC framework with the optimized individual violation rates, and then we propose a novel evolutionary algorithm to solve the intractable problem. Section V presents a statistical analysis of empirical solar forecast errors and the rationale behind the three data-driven ambiguity sets. Section VI presents a case study to evaluate the model performance and test the solution robustness. Finally, section VII concludes the paper.

\section{\textcolor{black}{Centralized OPF for networked microgrids}}

\begin{figure}[h]
    \centering
    \includegraphics[width=3.3in]{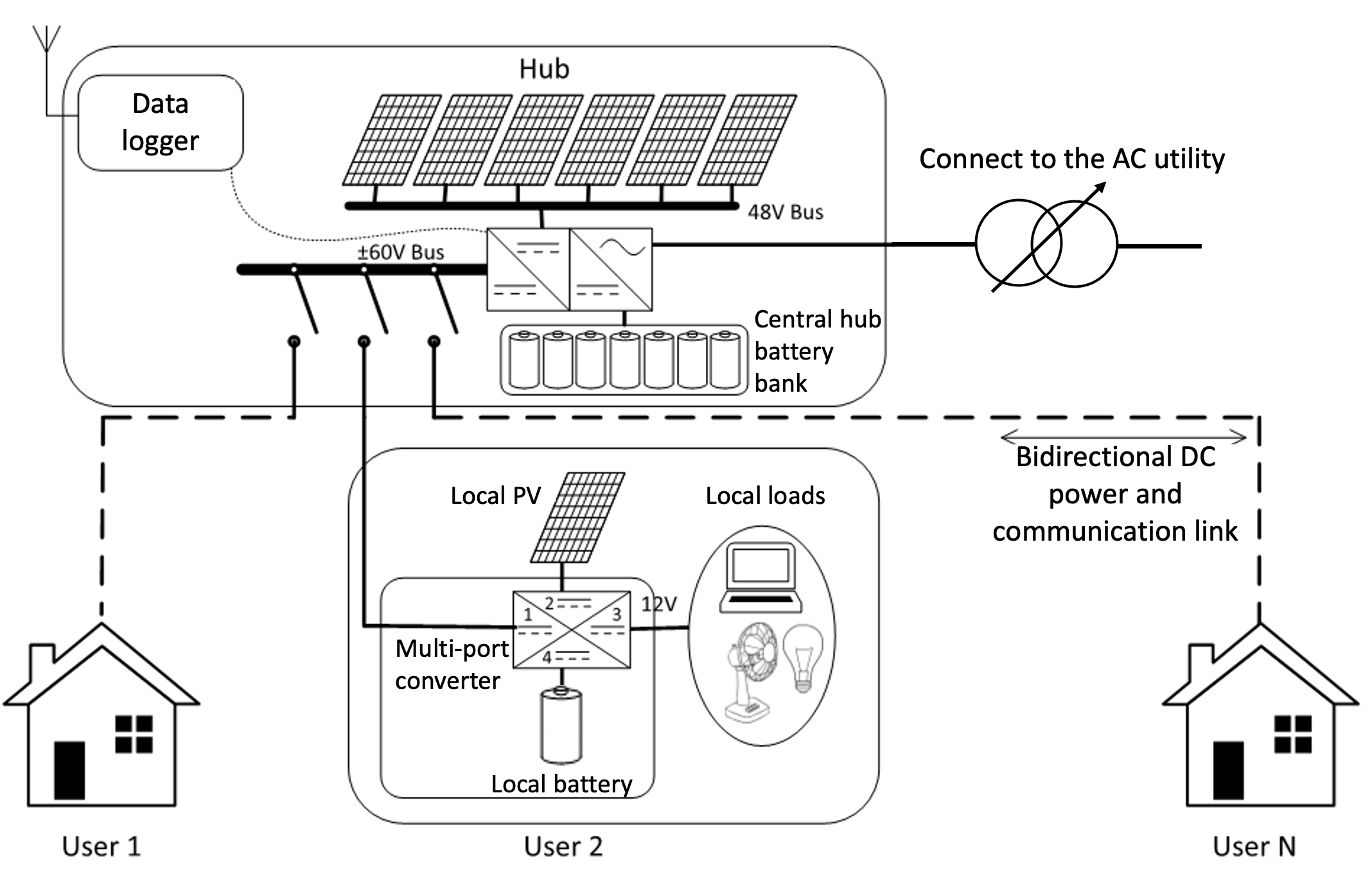}
    \caption{\textcolor{black}{Networked DC microgrid in the rural area \cite{Ding2020}}}
    \label{fig:3}
\end{figure}

Fig. \ref{fig:3} shows an example of a networked DC microgrid in a rural area. The microgrid has a main busbar connected to multiple households and the main grid via an inverter. The grid often experiences unpredictable power cuts. Each household at the end point has a bidirectional multi-port DC-DC converter connected to local PV panels, distributed energy storage (ES), and appliances. The centralized OPF is optimized in the receding horizon with 15-min time intervals and a one-day window. The formulation considers both the grid-connected and island mode simultaneously, allowing off-grid operation at any time step. The subsequent sections present the centralized OPF formulation with predetermined forecast errors.

\begin{subequations}

1) \textit{Preliminaries:}  
\textcolor{black}{As in Fig. \ref{fig:3}, we consider a networked microgrid with ${N_{B}}$ buses and  ${N_{L}}$ lines. The network buses are indexed by $b \in  \mathcal B $, and the network admittance matrix is denoted as $\textbf{Y}  \in \mathbb R^{N_B \times N_B }$. Distributed assets are located at different buses including energy storage, solar power generation, flexible and inflexible loads, indexed by $s \in \mathcal S $. Bold letters $\textbf{P}_{t} := \{P_{1, t}, P_{2, t},..., P_{n, t}\}$ represent decision variable vectors across distributed assets at time $t\in \mathcal T$. The positions of those distributed assets in the network are mapped by sparse matrices $\textbf{C}^{pv}  \in \mathbb R^{N_B \times N_{PV} }$, $\textbf{C}^{s} \in \mathbb R^{N_B \times N_{S}}$, $\textbf{C}^{l} \in \mathbb R^{N_B \times N_{D}} $. The multi-period centralized DC OPF has the time interval $ \Delta t $. For all constraints, $\forall t \in \mathcal T $ and $\forall b \in  \mathcal B $ hold unless otherwise specified. } 

2) \textit{Objective function:} The objective function is formulated,

\begin{equation}
\begin{aligned}
\mathcal J ={} &  \Delta t  {\sum^{T}_{t=0}\{ m_{u} (P_{t}^{g})^2 + m_{r} \sum_{s=0} ^{N_S} (R_{s, t}^{up})^2} \\
 &+m_{s} \sum_{s=0} ^{N_{PV}} (P_{s, t}^{pv} -  P_{s, t}^{pv'})^2 +m_{l} \sum_{s=0} ^{N_{D}}  (P_{s, t}^{l} -  P_{s, t}^{l'})^2 \\
 &+ m_{d}  \sum_{s=0} ^{N_{S}}(P_{s, t}^{d} +  P_{s, t}^{c}) \} 
 \label{obj_cost_d}
 \end{aligned}
\end{equation}

\textcolor{black}{The objective function includes five terms, namely, utility tariff, droop control provision cost, solar power curtailment penalty, load shedding penalty and battery degradation cost. Except for battery degradation costs, all costs are modeled as a quadratic function.} Cost factors $m_{s}$ and $m_{l}$ are associated with the solar power self-consumption and users' utility, as detailed in Section V.

2) \textit{SoC battery droop control}: Distributed ES in the network has two functions: shifting solar energy in time and providing P-V droop control to regulate bus voltages within the range. For the voltage regulation function, distributed ES reserves energy to address positive solar power forecast errors and provide upward voltage regulation. When solar power is overestimated, it leads to a power imbalance and low bus voltages. Then each ES will release the reserved energy according to its droop coefficient and bus voltage deviation, until a new power balance is achieved. When the solar power is under-estimated, the surplus solar power will be curtailed considering the curtailment penalty in (\ref{obj_cost_d}).
\begin{align}
       \textcolor{black}{\sum^{N_{S}}_{s=0} {d^{s}_{s, t} } = 1      \quad \forall d^{s}_{s, t} \in [0, 1] }       \label{droop_limit1} \\
        \textcolor{black}{R^{up}_{s, t}  \geq d^{s}_{s, t}  \sum_{s=0}^{N_{PV}} \tilde{P^{pv}_{s, t}}}  \label{droop control1} \\
         P^{d}_{s, t} + R^{up}_{s, t} \leq  \overline{P^{d}}        \label{dispower1} \\
         P^{c}_{s, t} - R^{up}_{s, t} \geq 0 \label{chpower1} \\
         \underline{SoC} \leq E^{s}_{s, t}  - R^{up}_{s, t} \Delta t   \label{disenergy1} \\
          E^{s}_{s, t} \leq \overline{SoC}  \label{chenergy1} \\
          \color {black}{{E^{s}_{s, t+1}} = E^{s}_{s, t} - \tfrac{{P^{d}_{s, t}}}{\eta_{dis}}\Delta t  + {P^{c}_{s, t}}\eta_{ch} \Delta t } \label{balance1}
\end{align}

\textcolor{black}{At any timestep $t$, all distributed ES in the network should deliver the P-V droop provision to address exactly the total amount of solar power forecast errors (\ref{droop_limit1}), while each distributed ES is coordinated to deliver a fraction known as the droop coefficient $d^{s}_{s, t}$ in (\ref{droop control1}). Droop provision is constrained by the battery power output limits (\ref{dispower1}) - (\ref{chpower1}), energy constraint (\ref{disenergy1})  - (\ref{chenergy1}) and energy balance considering the round-trip efficiency (\ref{balance1}).}

3) \textit{Power flow and balance:} We categorize users' loads into flexible and inflexible loads, and flexible loads can be curtailed during a blackout. 
\begin{align}
                   {P^{l'}_{s, t}} = {P^{l}_{s, t}} \label{load} \\
                    0 \leq {P^{pv'}_{s, t}} \leq {P^{pv}_{s, t}} \label{solar curtailment} \\
                     \langle P^{g}_{t},  P^{pv'}_{s, t}, P^{d}_{s, t},  P^{c}_{s, t},  {P^{l'}_{s, t}} \rangle \geq 0   \label{power flow} \\
                  \textcolor{black}{P^{g}_{t} +   \sum_{s=0}^{N_{PV}} {P^{pv'}_{s, t}} +   \sum_{s=0}^{N_{S}} ({P^{d}_{s, t} - P^{c}_{s, t}}) = \sum_{s=0}^{N_D} {P^{l'}_{s, t}}} \label{power balance}
\end{align}

At the grid-connected mode, the system cannot curtail any load (\ref{load}) but the solar power curtailment is allowed (\ref{solar curtailment}). All stacked decision variables in (\ref{power flow}) should be greater than zero. The power balance is guaranteed by (\ref{power balance}).
 
Considering distributed ES with the droop provision, local solar power generations with forecast errors and loads, the injected power at each bus in the network is formulated as,

\begin{equation} 
\label{injected1}
 \quad \textbf{P}^{inj}_{t}= \textbf{C}^{g} \textbf{P}^{g}_t+ \textbf{C}^{s} (\textbf{P}^{d}_t - \textbf{P}^{c}_t+ \textbf{R}^{up}_t) + \textbf{C}^{pv}  (\textbf{P}^{pv'}_t - \tilde{\textbf{P}^{pv}_t} )- \textbf{C}^{l} \textbf{P}^{l'}_t
 \end{equation}

The voltage at each bus depends on the power injected at that bus and power flow between all neighboring buses,

\begin{align}
        \textbf{P}^{inj} _{t} =  \textbf{diag}(\textbf{v}) \textbf{I} = \textbf{diag}(\textbf{v}) \textbf{Yv}  \label{voltage} \\
        \textbf{P}^{inj} _{t} = \textbf{diag}(\textbf{V}_{o}) \textbf{Y} \textbf{\textbf{v}} + \textbf{diag}(\textbf{I}_{o}) (\textbf{v} - \textbf{V}_{o})   \label{voltage_linear} \\
         \quad \underline{v} \leq \textbf{v} \leq \overline {v}     \label{voltage_limit} 
\end{align}

To deal with the non-convex constraint (\ref{voltage}), ref. \cite{Morstyn2016} uses \textit{Taylor series expansion} for linearization and validates its high fidelity. Based on this approach, constraint (\ref{voltage}) is linearized around the operating point $(\textbf{V}_{o}, \textbf{I}_{o})$ as (\ref{voltage_linear}). Bus voltages are regulated within a certain range to ensure power quality (\ref{voltage_limit}).

4) \textit{Island mode:} Designed to operate during a blackout, the microgrid can island at any timestep. The grid-connected and islanding schedules of the microgrid are solved simultaneously, as two scenarios of one problem. The optimization problem in the islanding scenario is identical to the grid-connected mode, except for the constraints pertaining to load curtailment and utility supply. 
\begin{align}
                   {P^{g}_{t}} = 0 \quad \quad t \in [t_{o}, t_{o}+H] \label{blackout} \\ 
                  {P^{cl}_{s, t}}\leq {P^{l'}_{s, t}} \leq {P^{l}_{s, t}} \quad \quad t \in [t_{o}, t_{o}+H] \label{load curtailment}
\end{align}

When the utility power is available, battery SoC of these two scenarios should be the same. When the blackout happens (\ref{blackout}) during $[t_{o}, t_{o}+H]$, the microgrid is only required to supply inflexible loads (\ref{load curtailment}) during islanding, and the battery control reference follows the optimization result of the islanding scenario.

\end{subequations}

\section{DR-JCC framework with the optimized individual violation rates}

\begin{figure}[h]
    \centering
    \includegraphics[width=2.7in]{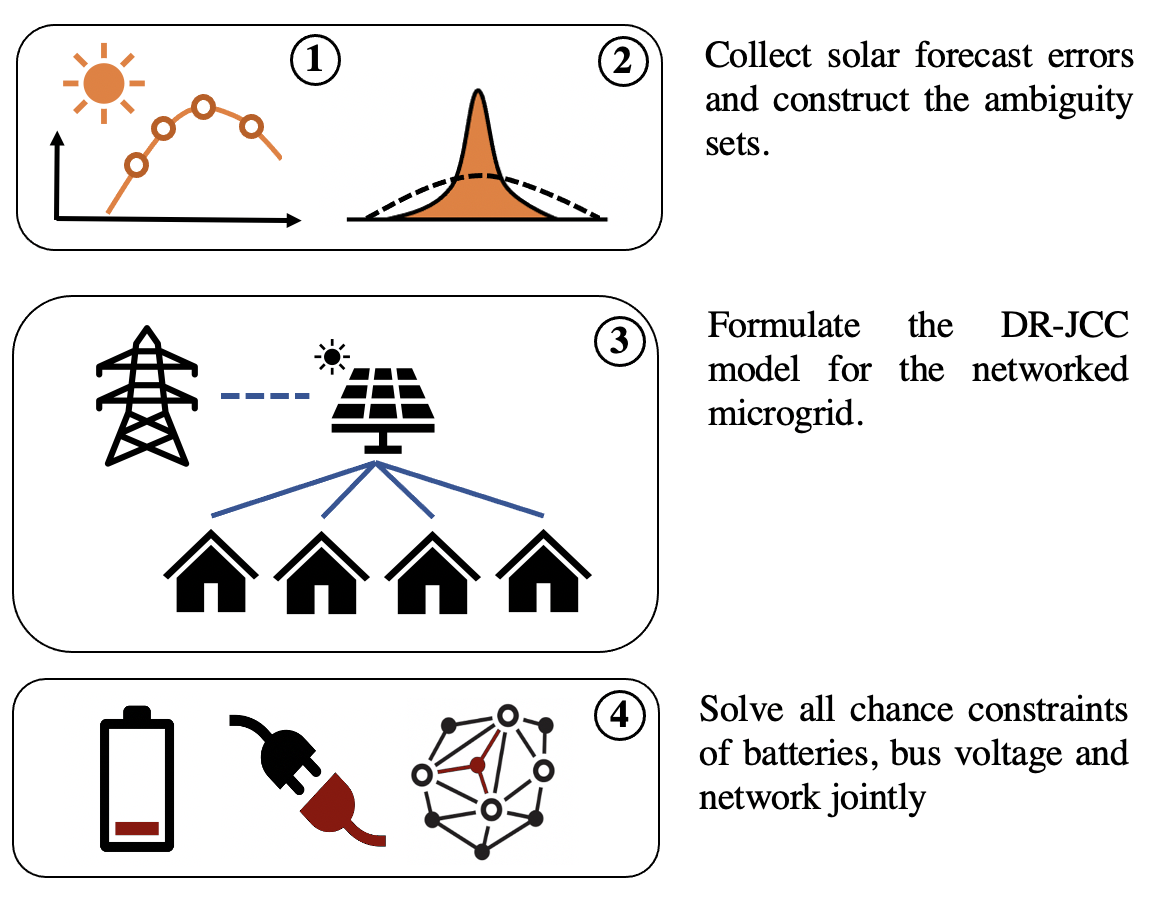}
    \caption{\textcolor{black}{Schematic diagram for formulating and solving the JCC model for the networked microgrid}}
    \label{schematic}
\end{figure}

\textcolor{black}{The centralized OPF formulation in section III has an underlying assumption of predetermined solar power forecast errors, $\tilde{P^{pv}_{s, t}}$, }while these errors follow an uncertain distribution in practice. We thus introduce the DR-JCC formulation to integrate uncertainty and secure solution robustness for uncertainty distributions. Introducing the JCC formulation is essential because the microgrid is reliable \textit{if and only if} all individual CCs are satisfied simultaneously. As shown in Fig. \ref{schematic}, the process for formulating and solving the DR-JCC problem includes four steps.

As shown in Fig. \ref{schematic}, firstly, the solar forecast error samples are collected. We then summarize the family of error distributions and build ambiguity sets. Given the shape and moment assumptions of ambiguity sets, the DR-JCC problem is formulated considering the \textit{joint} risk of the battery power capacity, bus voltage and network violation. The problem is solved by decomposing and recasting into the SOCP formulation. However, the JCC with the optimized violation rates is an intractable problem. Thus, we propose a novel evolutionary algorithm to solve this problem. 

\subsection{DR-JCC formulation}
We first present the DR-JCC formulation considering uncertain solar power forecast error distributions. All six constraints (\ref{droop control1}) - (\ref{disenergy1}) and (\ref{voltage_limit}) involving forecast errors $\tilde{P^{pv}_{s, t}}$ are reformulated in the DR-JCC fashion, while the other constraints (\ref{droop_limit1}), (\ref{chenergy1}) -  (\ref{voltage_linear}), (\ref{blackout}) and (\ref{load curtailment}) remain the same. The new problem formulation is given by, 

$
 \begin{aligned}
     \label{optimization}
 {}
& \min_{x}  \mathbb{E}_\mathbb{P}[\mathcal J(x)] \\
  & \textrm{s.t. }  \text{(\ref{droop_limit1}), (\ref{chenergy1}) -  (\ref{voltage_linear}), (\ref{blackout}), (\ref{load curtailment})} \\
    \end{aligned}
$

\begin{equation}
    \label{joint_cc}
 \begin{aligned}
 {}
&\inf_{\mathbb{P} \in \mathcal P (\mu, \sigma)} \mathbb{P}( \textstyle \bigcap^{N_c}_{i=0}({A}_i (x) \zeta \leq{b}_i (x) )) \geq 1-\epsilon_j
    \end{aligned}
\end{equation}

\begin{subequations}

\textcolor{black}{
where $ {A}_i (x) \zeta \leq {b}_i (x) =$
\begin{align}
         \{d^{s}_{s, t} \sum_{i=0}^{N_{PV}} \tilde{P^{pv}_{s, t}} \leq R^{up}_{s, t}   \label{droop control2} \\
         P^{d}_{s, t} + R^{up}_{s, t} \leq  \overline{P^{d}}        \label{dispower2} \\
         R^{up}_{s, t} \leq P^{c}_{s, t}  \label{chpower2} \\
         E^{s}_{s, t}  - R^{up}_{i, t} \Delta t  \geq   \underline{SoC} \label{energy2} \\
         \textbf{v}  \geq  \underline{v} \label{voltage lower2}\\
          \textbf{v} \leq \overline {v} \label{voltage upper2}\} 
\end{align}
}

\end{subequations}

The letter $x$ represents the decision variable vector (i.e. droop coefficient, grid power and distributed asset power outputs), and $\zeta$ represents the uncertainty variable vector following the distribution $\mathcal{P} $ (i.e. solar power forecast error $\tilde{P^{pv}}$). $N_c$ is the number of individual constraints included, while $ {A}_i (x)$ and ${b}_i (x)$ are affine functions about the decision variables. The DR-JCC inequality (\ref{joint_cc}) means that given all distributions in the ambiguity set $\mathcal P $ built upon moments $\mu$, $\sigma$, the violation rate of the JCC is less than $\epsilon_j $ even for the worst-case distribution. \textcolor{black}{This JCC problem can be decomposed into the SCC problem with individual violation rates. Constraint (\ref{joint_cc}) is transformed to constraint (\ref{icc1}) based on the set operation properties of supremum and infimum. Then the JCC (\ref{icc1}) can be decomposed into $N_c$ single chance constraints based on \textit{Boole's inequality} \cite{McKinsey1950}.}

 \begin{align}
 \color{black}
\sup_{\mathbb{P} \in \mathcal P(\mu, \sigma)} \mathbb{P}(\textstyle \bigcup^{N_c}_{i=0}( {A}_i (x) \zeta > {b}_i (x) ))  \leq \epsilon_j   \label{icc1}\\
 \color{black}
\xRightarrow {} \sup_{\mathbb{P} \in \mathcal P (\mu, \sigma)} \textstyle \sum ^{N_c}_{i=0} \mathbb{P}( {A}_i (x) \zeta >  {b}_i (x) )  \leq \epsilon_j  \label{icc2}
\end{align}

Thus, given the violation rate of the single CC $\epsilon_i$, the following inequality holds.

\begin{equation} 
\label{oba}
 \sum_{i=0} ^{N_c} \epsilon_i \leq \epsilon_j 
\end{equation}

\subsection{Ambiguity set construction and SOCP formulation}

The solar forecast error distributions typically show strong unimodality or symmetry\cite{Li2019}. This paper uses moments and distribution structure assumptions to construct ambiguity sets. For a time interval $t \in \mathcal {T}$, we have $N_s$ error samples $\zeta_{n, t}$. We can compute the empirical mean $\mu_t$ and covariance $\sigma_t$ of the error distribution in this time interval, 

\begin{equation}
    \label{mean}
\mu_t =  \frac{1}{N_s} \sum_{n=0}^{N_s} \zeta_{n, t}
\end{equation}

\begin{equation}
    \label{cov}
\sigma_t^2 =   \frac{1}{N_s-1} \sum_{n=0}^{N_s} (\zeta_{n, t} - \mu_t) (\zeta_{n, t}- \mu_t)^\intercal
\end{equation}

Given the solar power correlation between two buses $a, b$ at time $t$ denoted as $\gamma_{a, b, t}$, the mean vector $\mu_{pv}$ and covariance matrix $ \Sigma_{pv}$ of forecast errors in the network are, 

\begin{equation}
    \label{mean_tot}
       \mu_{pv} :=  [\mu_{0, t}, \cdots, \mu_{b, t}] \quad
\end{equation}

\begin{equation}
s_{a, b} = 
\begin{cases} 
\sigma_{a, t}^2 &\mbox{if } a = b \\
\gamma_{a,b,t}\sigma_{a, t}\sigma_{b, t}& \mbox{if } a \neq b \end{cases}
\quad( \forall s_{a, b}  \in \textbf{S} \in \mathbb R^{N_B \times N_B })
\end{equation}

\begin{equation}
    \label{cov_matrix}
      \Sigma_{pv} := \textbf{S}-\mu_{pv} \mu_{pv}^\intercal
\end{equation}

Depending on the certainty of the computed moments, one can build the ambiguity set and recast the CC model into either the SOCP or SDP formulation. The former considers no estimation errors for the computed moments, while the latter considers the confidence level of the moment estimation. A detailed demonstration can be consulted in \cite{Delage2010a}. In this paper, we prepare a sufficient number of samples and adopt the SOCP formulation assuming the computed moments are exactly true moments of unknown distributions. We construct three ambiguity sets as $\mathcal D^1_{\zeta}$, $\mathcal D^2_{\zeta} $ and $\mathcal D^3_{\zeta}$. We drop indexes $t, pv$ for conciseness.

\textbf{Ambiguity set 1}: (Unimodal, centred at the mean and mode zero.)

\DeclarePairedDelimiter\set{\{}{\}}
\newcommand\numberset[1]{\mathbf{#1}}
\newcommand\Real{\numberset{R}}

\begin{align*} 
\mathcal D^1_{\zeta} := \set*{\mathbb P\in \mathcal P^{\alpha} : \begin{matrix} \mathbb E[\zeta] = \mu,  \mathbb E[(\zeta-\mu)^2] = \sigma^2 \\
 \mathcal M[\zeta] = \mu = 0 \end{matrix}}
\end{align*}

where $\mathcal P^{\alpha}$ denotes all unimodal distributions on $\mathbb R^{n}$, and $\mathcal M[\zeta]$ is the mode of distributions. 

\textbf{Ambiguity set 2}: (Symmetric, centred at mean zero.)

\begin{equation}
\label{ambiguity set_2}
\mathcal D^2_{\zeta} := \set*{\mathbb P\in \mathcal P : \begin{matrix} \mathbb E[\zeta] = \mu, \mathbb E[(\zeta-\mu)^2] = \sigma^2 \\
\mathbb P [\zeta] = \mathbb P [-\zeta]  = 0 \end{matrix}}
\end{equation}
 
\textbf{Ambiguity set 3}: (Unimodal and symmetric, centred at mean and mode zero.)

\begin{equation}
\label{ambiguity set_2}
\mathcal D^3_{\zeta} := \set*{\mathbb P\in \mathcal D^1_{\zeta} \cap  \mathcal D^2_{\zeta}}
\end{equation}

Based on the proof in \cite{Calafiore2006}, a single DR-CC in (\ref{icc2}) can be recast into a SOCP constraint given by, 

\begin{equation} 
\label{socp}
\lambda (\epsilon_i)  ||\Sigma^{1/2}  A_i (x) ||_2 \leq b_i(x) - {\mu}^\intercal A_i (x)
\end{equation}

where $\lambda (\epsilon_i)$ depends on the specific ambiguity set. For $\mathcal D^1_{\zeta}$, $\mathcal D^2_{\zeta}$ and $\mathcal D^3_{\zeta}$, their functions $\lambda_1 (\epsilon_i)$, $\lambda_2 (\epsilon_i)$ and $\lambda_3 (\epsilon_i)$ are,  

\begin{equation} 
\label{rho_unimodal_1}
 \lambda_1 (\epsilon_i)  :=  \frac{2}{3}\sqrt{\frac{1}{\epsilon_i}} \quad \forall \epsilon_i \in (0, \frac{1}{3})
 \end{equation}

 \begin{equation} 
\label{rho_unimodal_2}
\lambda_2 (\epsilon_i)  :=  \sqrt{\frac{1}{2\epsilon_i}} \quad \forall \epsilon_i \in (0, \frac{1}{2})
 \end{equation}
 
  \begin{equation} 
\label{rho_unimodal_3}
 \lambda_2 (\epsilon_i)  :=  \sqrt{\frac{2}{9\epsilon_i}} \quad \forall \epsilon_i \in (0, \frac{1}{6})
 \end{equation}
 
Equations (\ref{rho_unimodal_1}) - (\ref{rho_unimodal_2}) are proven in \cite{Popescu2005} and \cite{VanParys2016} based on \textit{Gauss's inequality} and \textit{Chebyshev's inequality}. Equation (\ref{rho_unimodal_3}) is proven in ref. \cite{Ding2021b}.

\subsection{\textcolor{black}{Approximation of individual violation rates}}

Since we only know a joint violation rate $\epsilon_{j}$, single violation rates $\epsilon_{i}$ need to be computed. We choose \textit{Bonferroni Approximation} as the baseline case. It assumes all individual violation rates to be equal and the sum of individual violation rates is the exact joint violation rate.

\begin{equation} 
\label{ba}
 \epsilon_i := \frac{\epsilon_j}{N_c}
\end{equation}

Nevertheless, the \textit{Bonferroni Approximation} leads to a very conservative solution, as this approximation assumes all constraints have the same chance to be violated. We therefore propose an improved method using the \textit{Optimized Bonferroni Approximation}. It considers the individual violation rate $\epsilon_i$ as a variable rather than a fixed priori, which is solved simultaneously with the original optimization problem \cite{Xie2019}.

 \begin{align}
        \textbf{H} :=  -{\textbf{C}^{s}\textbf{d}^s_t + \textbf{C}^{pv}}  \label{h} \\
        \textbf{G} :=  \textbf{diag}(\textbf{V}_o)\textbf{Y}+ \textbf{diag}(\textbf{I}_o) \\
         \lambda (\epsilon_i)  ||\textbf{d}^{s}_{t} \textbf{1} \Sigma^{1/2}_{pv, t} ||_2 \leq \textbf{R}^{up}_{t} - \mu_{pv, t}^\intercal \textbf{1} \textbf{d}^{s}_{t}   \label{so1} \\
        \lambda (\epsilon_i)  ||\textbf{d}^{s}_{t} \textbf{1} \Sigma^{1/2}_{pv, t}  ||_2 \leq  \overline{\textbf{P}^{d}}- \textbf{P}^{d}_{t}-\mu_{pv, t}^\intercal \textbf{1} \textbf{d}^{s}_{t}   \label{so2} \\
        \lambda (\epsilon_i) ||\textbf{d}^{s}_{t} \textbf{1} \Sigma^{1/2}_{pv, t} ||_2 \leq \overline{\textbf{P}^{c}}- \mu_{pv, t}^\intercal \textbf{1} \textbf{d}^{s}_{t}   \label{so3}
\end{align}

\begin{equation} 
\label{so4}
\lambda (\epsilon_i)  ||-\textbf{d}^{s}_{t} \textbf{1} \Delta t \Sigma^{1/2}_{pv, t}  ||_2 \leq  {\textbf{E}^{s}_{t}}- \underline{\textbf{SoC}} - (- \mu_{pv, t}^\intercal  \textbf{1}  \Delta t \textbf{d}^{s}_{t})
\end{equation}
         
\begin{equation} 
\label{so5}
  \lambda(\epsilon_i)  || \textbf{G}^{-1} \textbf{H} \Sigma^{1/2}_{pv, t} ||_2 \leq  {\overline{\textbf{v}}} - \textbf{G}^{-1}\{ \textbf{diag}({\textbf{I}_{o}})\textbf{V}_{o} + \textbf{P}^{inj}_{t}\}
 \end{equation}

\begin{equation} 
\label{so6}
  \lambda(\epsilon_i)  ||\textbf{G}^{-1} \textbf{H}  \Sigma^{1/2}_{pv, t} ||_2 \leq -{\underline{\textbf{v}}} +  \textbf{G}^{-1} \{\textbf{diag}({\textbf{I}_{o}})\textbf{V}_{o} +  \textbf{P}^{inj}_{t}\}
 \end{equation}

Based on the \textit{Optimized Bonferroni Approximation} and the reformulation in (\ref{socp}) - (\ref{rho_unimodal_3}), the DR-JCC inequality (\ref{joint_cc}) consisting of six individual CCs (\ref{droop control2}) - (\ref{voltage upper2}) is recast into individual SOCP constraints (\ref{so1}) - (\ref{so6}), with a new variable $\epsilon_i$ and constraint (\ref{oba}). \textbf{1} represents the unit vector with the dimension of buses, and $\mu_{pv, t}^\intercal \textbf{1}$ thus is the sum of total forecast error across all buses. However, introducing the variable $\epsilon_i$ destroys the convexity of this problem. One can observe that each constraint has a multiplication of variables. The problem has been proven as strongly NP-hard \cite{Xie2019}.
 
 \subsection{Evolutionary algorithm for the JCC problem}

\textcolor{black}{To tackle the aforementioned NP-hard problem, we propose an evolutionary algorithm  to approximate the solution of the intractable JCC problem, including the optimized individual violation rate for each CC and total operation cost. In the subfield of meta-heuristic optimization, the evolutionary algorithm is a bio-inspired algorithm analogous to the natural evolution process \cite{DarioFloreano2009}. The essence of an evolutionary approach to solve a problem is to equate possible solutions to individuals in a population, and to introduce a notion of fitness on the basis of solution quality \cite{Eiben2015}.}

\begin{algorithm}[h]
\SetAlgoLined
\textit{Initialise} the population with random individual solutions; 

\textit{Evaluate} each individual solution; 

\While{termination condition is not satisfied}{
    Perform \textit{competitive selection}; \\
    Apply \textit {pair, breed and mutation} procedures; \\
    Evaluate the new pool of individual solutions; \\
    Apply \textit{replacement} to form the new
 population; \\
\textit{Find} current best solution; 
    }
    \textit{Output} overall best solution; 
 \caption{Population-based evolutionary algorithm}
\end{algorithm}

\textcolor{black}{Algorithm 1 provides the pseudo-code of the proposed population-based evolutionary algorithm. First, a group of $N_p$ individuals $\textbf{P}$ as the first-generation population is created,}

\begin{equation} 
\label{individual}
 \textbf{P}:= [\epsilon_{0}, ...., \epsilon_{i}] \quad  \forall i \in [0, N_c],  \forall \epsilon_i \in [\underline{\epsilon}, \overline{\epsilon}]
 \end{equation}
 
Each individual $\textbf{P}$ has six parameters, which are the optimized variables for each CC violation rate (\ref{individual}). The sum of individuals' parameters should satisfy \textit{Boole's inequality} (\ref{oba}), thus the upper bound of parameters is the joint violation rate $\epsilon_j$. We use the voltage violation rate of a feeder in reality (e.g. 0.1\% \cite{Palmintier2016}) to set the lower bound of parameters. 

\begin{equation} 
\label{objective}
\mathcal{F}(\textbf{P}):=\mathbb{E}_\mathbb{P}[\mathcal J(\textbf{P}, x)]
 \end{equation}
 
Each individual is evaluated by the fitness value, defined as the objective function value of the JCC problem (\ref{objective}), where $x$ is the decision variable in the original optimization problem to be solved together. The evolution of the population is conducted iteratively based on competitive selection. For each iteration, only the first half of individuals with lower fitness values are selected as the \textit{elite} for the next generation. 

 \begin{equation} 
\label{generation_1}
\epsilon^s_{i} := \frac{\epsilon^{cm}_{i} + \epsilon^{cn}_{i}}{2} \quad \forall \epsilon^{cm}_{i} \in \textbf{P}_{m}, \forall \epsilon^{cn}_{i} \in \textbf{P}_{n}
 \end{equation}
 
  \begin{equation} 
\label{mutation_1}
\epsilon^{s'}_{i} := \epsilon^{s}_{i} + \max\{\theta, 0\} \quad  \theta \sim \mathcal{N}(0,\,\sigma^{2})
 \end{equation}
 
The \textit{elite} pairs with each other and generates offspring. Specifically, the offspring is generated by taking the average value of parents'  parameters (\ref{generation_1}). The mutation of offspring is necessary, otherwise the solution might be trapped into a local minimum. That is to add a random number drawn from a normal distribution to six offspring' parameters (\ref{mutation_1}). The parameters of the mutated offspring should be normalized, so that the sum is always equal to the joint violation rate. The mutation only happens when all parents are different, otherwise the fitness values over generations are hard to converge. 
 
The evolutionary iteration will stop until any of these two termination criteria is reached, the maximum number of iterations (i.e. $N_{it}$ = 10), and the ratio to measure the solution convergence, defined as the ratio of the maximum. The average value of fitness functions is as below.

\begin{equation} 
\label{termination_1}
r := \frac{\max\{\mathcal{F}(\textbf{P}_0),...,\mathcal{F}(\textbf{P}_n) \}}{\frac{1}{N_p}\sum^{N_p}_{n=0}\mathcal{F}(\textbf{P}_n)} - 1\leq r_{thr}
 \end{equation}

The algorithm will stop when the ratio (\ref{termination_1}) is lower than the threshold value, $r_{thr}$. \textcolor{black}{Much empirical evidence such as \cite{Eiben2015}, \cite{Thakur2014} and \cite{Coello2004} shows that if the evolutionary algorithm, as one of the global optimization methods, is repeated many times with the random initial guesses in the first generation and still obtains the same solution, this solution is considered as an acceptable approximation of the global optimum.} 

\textcolor{black}{This method can be readily transferable from one problem to another, as only two parts of the algorithm are problem-dependent, the initial values of the first generation (i.e., the initial guess of individual violation rates) and fitness function (i.e., the objective function).   Moreover, one can adjust the convergence ratio (\ref{termination_1}) of the algorithm to decide the trade-off between the fidelity of the optimum approximation and the computation time. However, if a problem has a very small joint violation rate divided by a great number of chance constraints, some of the initial guesses for the optimized violation rates could result in infeasible solutions. A successful evolution process would require a large population and significant computation efforts.} 

\section{Data-driven solar power forecast}
 
We choose a light gradient boosting machine in ref. \cite{DataStudyGroupTeam2020} to predict solar power. This paper does not consider the demand uncertainty since its forecast errors are generally much smaller than solar power forecast errors for a microgrid with a high solar self-consumption. The dataset used is two-year 15-min weather measurements in Gitaru dam, Kenya, including the solar irradiance, air temperature and wind speed for the prediction features. We set the prediction horizon to be one day. The prediction is updated every time interval in the receding horizon for a whole year.

\begin{figure}[h]
    \centering
    \includegraphics[width=3in]{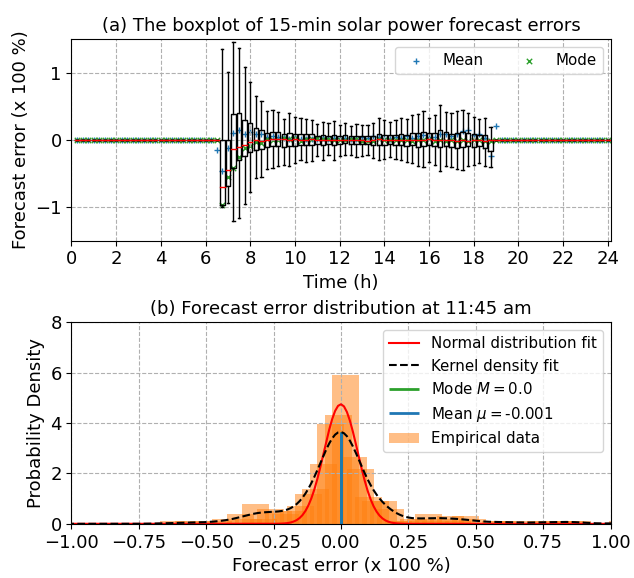}
    \caption{Solar power forecast error distributions in a year (a) the boxplot of 15-min solar power forecast error (modes and means labeled with black and green markers) (b) Forecast error distribution at 11:45 am (with non-parametric kernel density estimation in black dash line and parametric Gaussian fitting in red solid line)}
    \label{histogram}
\end{figure}

We calculate solar power forecast errors all over the year, aggregate them based on time intervals and label the mean and mode, as in Fig. \ref{histogram} (a). For each time interval with the daylight, there are 365 error samples from each day in a year. The maximum value is around 150\% in the early morning when the solar power is too small (around 1W) to be accurately predicted. One can also observe several phenomena in Fig. \ref{histogram}. First, those error distributions in each time interval are highly symmetric centered at their means. Second, both the mean and mode are around zero and fairly close except for early morning and late afternoon when the solar irradiance is very low. Third, the distribution is unimodal but not necessarily a normal distribution as shown in Fig. \ref{histogram} (b). 

\section{Case Study}

The section presents a case study of a networked DC microgrid in the rural area, Kenya. This microgrid has 10 households connected to the main busbar radially, and each has local PV panels and 60Wh, 20W home batteries. Their power outputs are considered to be independent. Households' load profiles are constructed using the weather data based on our previous work \cite{Ding2020}. Specifically, inflexible (i.e. light) and flexible (i.e. fan and phone charger) loads are considered, and the utility for using them is modeled as a quadratic function of power, temperature and solar irradiance based on welfare economics \cite{Li2011a}. The parameter $m_l$ is a coefficient of the utility function to reflect users' comfort  to achieve smart load shedding. For example, the user's utility when using lights is inversely proportional to the solar irradiance, meaning that the more economic utility that users can get from lights during the low solar irradiance. 

\begin{table}[!h]
\caption{Case study parameters}
\label{table_2}
\centering
\begin{tabular}{llcccc}
\toprule
$m_u$ &  \$0.023/(kWh)$^{2}$ & $\eta_{dis}, \eta_{ch}$ & 0.95 &$N_p$& 6 \\
$m_s$  & \$1.00/(kWh)$^{2}$  & $\underline{SoC}, \overline{SoC}$ &0.2, 1&$N_{it}$& 10\\
$m_v$  &  \$0.23/(kWh)$^{2}$ &$\underline{v_{b}}, \overline{v_{b}}$  & 0.95, 1.05p.u. &$r_{thr}$& 2\%\\
$m_d$   &  \$0.27/kWh &$R_{line}$  & 8$\Omega$/km &$\sigma_m$& 0.1\\
\bottomrule
\end{tabular}
\end{table}

The power capacity of PV panels and line length from main bus to households are drawn from uniform distributions $\mathcal{U}_1 (20, 40)$ [W] and $\mathcal{U}_2 (50, 200)$ [m]. Table \ref{table_2} shows the model parameters in the first two columns and the evolutionary algorithm parameters in the third column. The model parameters such as line resistance are from the manufacturer information \cite{RS}, and $m_s$ is set to be large to encourage self-consumption. The evolutionary algorithm parameters are optimized from empirical experiments. For example, we increase the number of individuals $N_p$ by two each time and observe the diminishing marginal improvements on computation time and results, until there is no significant improvements on results. The model framework is built using the CVXPY package \cite{diamond2016cvxpy} in Python, and run on an Apple iMac with a processor of 3.1GHz Intel Core i5 and a memory module of 8 GB 2133 MHz LPDDR3.

\subsection{Solving DR-JCC problem with evolutionary algorithm}

We first solve the DR-JCC model with different joint violation rates using the proposed evolutionary algorithm, and compare results with the baseline, the \textit{Bonferroni Approximation}. \textcolor{black}{In each experiment with one joint violation rate, we run the evolutionary algorithm for 10 times and the solutions converge to the same optimum. Those solutions are thus considered as an acceptable approximation of the global optimum. }

\begin{figure}[!h]
    \centering
    \includegraphics[width=3in]{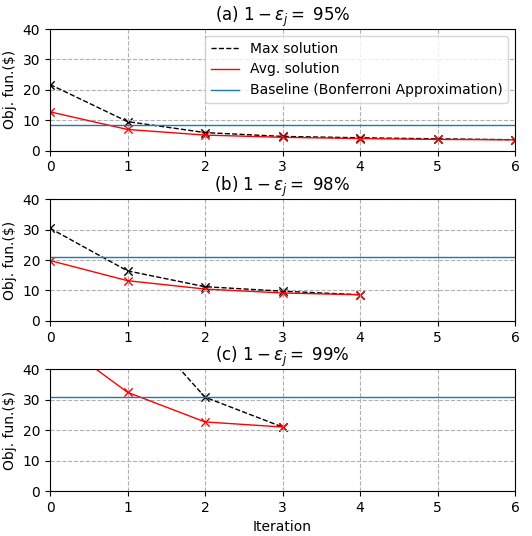}
        \caption{Computation processes for DR-JCC under three joint violation rates}
    \label{result5}
\end{figure}

The computation processes for three joint violation rates (i.e. 0.05, 0.02, 0.01) with the unimodal ambiguity set $\mathcal D^1_{\zeta}$ are shown in Fig. \ref{result5}. First, in the higher system reliability (i.e. $1- \epsilon_{j}$) cases, solutions converge faster. This is because the numerical range of individuals' parameters to explore becomes smaller. Second, the substantial reduction of the objective function is often achieved in the first two or three iterations. This means one can decide the trade-off between computation time and solution conservativeness by changing the termination condition such as the ratio $r_{thr}$. 

\begin{table}[!h]
\caption{Summary of results and computation times}
\label{solving_jcc}
\begin{tabular}{@{}ccccccc@{}}
\toprule
\textbf{\begin{tabular}[c]{@{}c@{}}Joint vio. rates \end{tabular}}      & \multicolumn{2}{c}{\textbf{0.05}} & \multicolumn{2}{c}{\textbf{0.02}} & \multicolumn{2}{c}{\textbf{0.01}} \\ \midrule
\textbf{Methods}                                                            & BL.                     & PPSD.  & BL.                     &  PPSD.  & BL.                     & PPSD. \\ \midrule
$\epsilon_{1}$                                                           & \multirow{3}{*}{0.0083} & 0.023    & \multirow{3}{*}{0.0033} & 0.008    & \multirow{3}{*}{0.0016} & 0.003    \\
$\epsilon_{2}$                                                               &                         & 0.022    &                         & 0.008    &                         & 0.003    \\ 
$\epsilon_{3} - \epsilon_{6}$                                                             &                         & 0.001    &                         & 0.001    &                         & 0.001    \\ 
 \midrule
\textbf{Time (s)}                                                        & 11.02                    & 464.02     & 11.69                   & 351.82     & 12.55                    &  269.14    \\ \midrule
\textbf{\begin{tabular}[c]{@{}c@{}}Obj.  func. (\$)\end{tabular}} &8.54                     & 3.54    & 21.54                      & 8.54       & 30.86                     & 21.06      \\ \bottomrule
\end{tabular}
    \begin{tablenotes}
      \footnotesize
      \item BL.: Baseline; PPSD.: Proposed
    \end{tablenotes}
\end{table}

Detailed results and computation times are listed in Table \ref{solving_jcc}. Significant cost reductions are achieved in all runs compared with the baseline method. For three cases, the reduction is 58.50\%, 60.35\% and 31.75\% respectively. \textcolor{black}{Parameters $\epsilon_{1}$ - $\epsilon_{6}$ are the optimized individual violation rates of chance constraints (\ref{so1}) - (\ref{so6}). These constraints are the voltage droop regulation, battery power discharging and charging limits, battery energy limit, voltage upper and lower regulations. Among all six violation rates, parameters $\epsilon_{1}$ and $\epsilon_{2}$ are optimized to have the higher values, while parameters $\epsilon_{3}$ - $\epsilon_{6}$ are optimized to have the lower values (i.e. 0.001). To check how these parameters change with the power flow conditions, we change the line resistance $R_{line}$ from 8 to 12 $\Omega$/km and repeat experiments. The operation cost slightly increases but the value of parameters remains unchanged. }

The results indicate the first two constraints are the most critical to the operation costs. They are droop regulation and battery discharging power limit. This means the operation cost is mainly determined by the droop provision and solar power uncertainty, as we set the high value for cost coefficients for energy reserve and solar curtailment. On the other hand, tightening the last four constraints, battery power charging limit, bus voltage and battery energy limits will not increase the cost significantly, even with a high line resistance. This is because the objective function does not include the monetary term for power loss pertaining to bus voltages.
  
\subsection{DR-JCC framework performance}

As this model framework aims to address renewable uncertainty in microgrids and tackle utility contingencies, we conduct simulations with a progressively smaller joint violation rate ranging from 0.2 to 0.01. The utility blackout is simulated with a duration ranging from one hour to a day.

\begin{figure}[!h]
    \centering
    \includegraphics[width=3.5in]{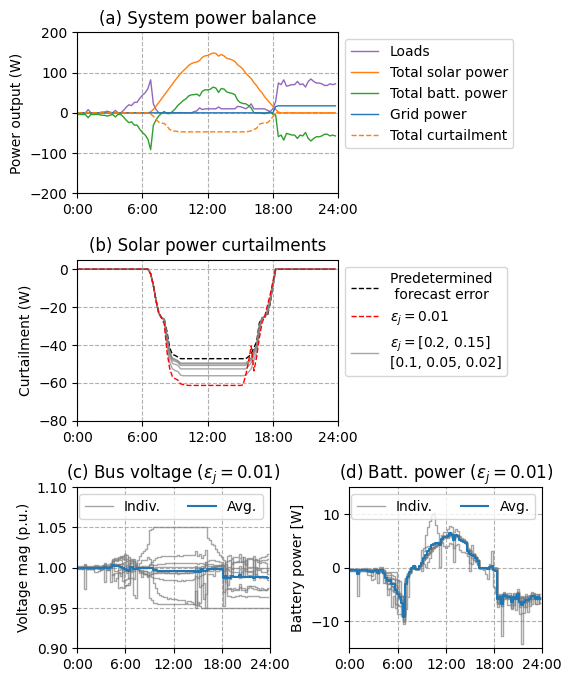}
        \caption{(a) System power balance, (b) solar power curtailment, (c) bus voltages when $\epsilon_{j} = 0.01$ and (d) batteries' power outputs when $\epsilon_{j} = 0.01$ in the DR-JCC framework. The average values are plotted in blue thick lines in (c) and (d) (with 50\% opacity to show individual values)}
    \label{deterministic}
\end{figure}

Fig. \ref{deterministic} showcases how the microgrid system addresses solar generation uncertainty under different joint violation rates $\epsilon_{j}$, including (a) overall power flow, (b) solar power curtailment, (c) bus voltages and (d) battery powers when $\epsilon_{j} = 0.01$. For battery power outputs, negative values represent the battery discharging. In the DR-JCC framework, the microgrid system curtails the increasing amount of solar power with a progressively tighter joint violation rate in Fig. \ref{deterministic} (b). Fig. \ref{deterministic} (c) and (d) show bus voltages and battery powers of individual households with the average value. The bus voltages are regulated within a range of $\pm 0.05 $ p.u..

\begin{figure}[!h]
    \centering
    \includegraphics[width=3.5in]{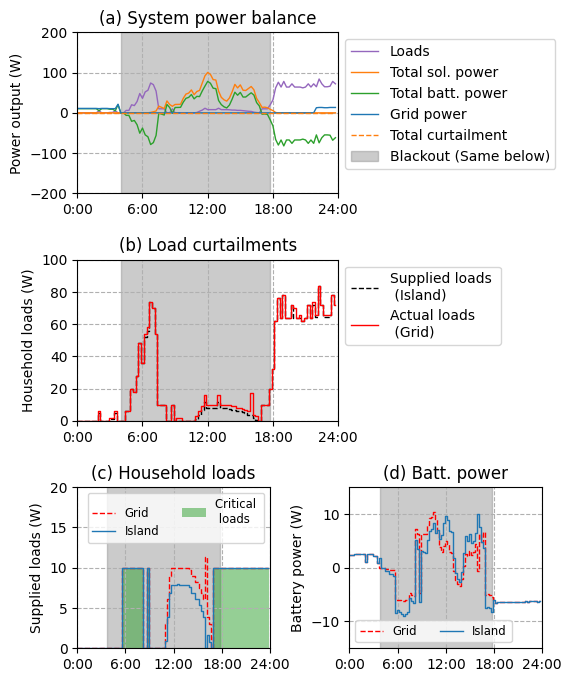}
        \caption{(a) System power balance, (b) load curtailments, (c) one household load supply and (d) one household battery power outputs when there is a blackout. In (b), (c) and (d), two scenarios are presented, the grid-connected one under the normal condition and the island one for blackout.}
    \label{system_blackout}
\end{figure}

The islanding operation of the microgrid is simulated for an overcast day with a blackout between 4 a.m. - 6 p.m. Fig. \ref{system_blackout} shows the simulation result. In Fig. \ref{system_blackout} (b), when a blackout happens, flexible loads are curtailed, mainly phone changer (i.e. load spike at 17 p.m.) and fan at noon. Battery control reference follows the islanding mode solution during the blackout, as blue solid lines in Fig. \ref{system_blackout} (d).  

\begin{figure}[!h]
    \centering
    \includegraphics[width=3in]{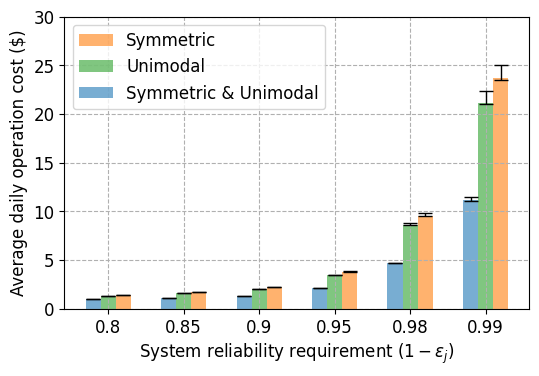}
        \caption{Daily operation costs with the three ambiguity sets (with the error bar to show the cost deviation due to the island hour from no blackout to a full-day blackout)}
    \label{summary}
\end{figure}

We compute daily operation costs under six different reliability requirements (i.e. $1-\epsilon_{j}$) and three ambiguity sets with the 12-hour islanding period. The error bars show the cost deviations due to the islanding time from no blackout to a full-day blackout. In summary, in Fig. \ref{summary}, the daily operation cost increases exponentially along with the system reliability requirements for all three ambiguity sets, and extending the islanding duration also increases the operation cost.

\subsection{Solution robustness and system reliability}

Following the performance demonstration, this section presents solution robustness tests for the proposed evolutionary algorithm and system reliability outcomes. We use forecast error samples excluding those used to construct the ambiguity set. \textcolor{black}{We fix the solution obtained and run the power flow under those forecast error samples, then count the case when all the constraints are met (i.e., no-violation case). The percentage of no-violation time intervals out of total time intervals in a day is defined as the daily reliability outcome.}

\begin{figure}[!h]
    \centering
    \includegraphics[width=3.5in]{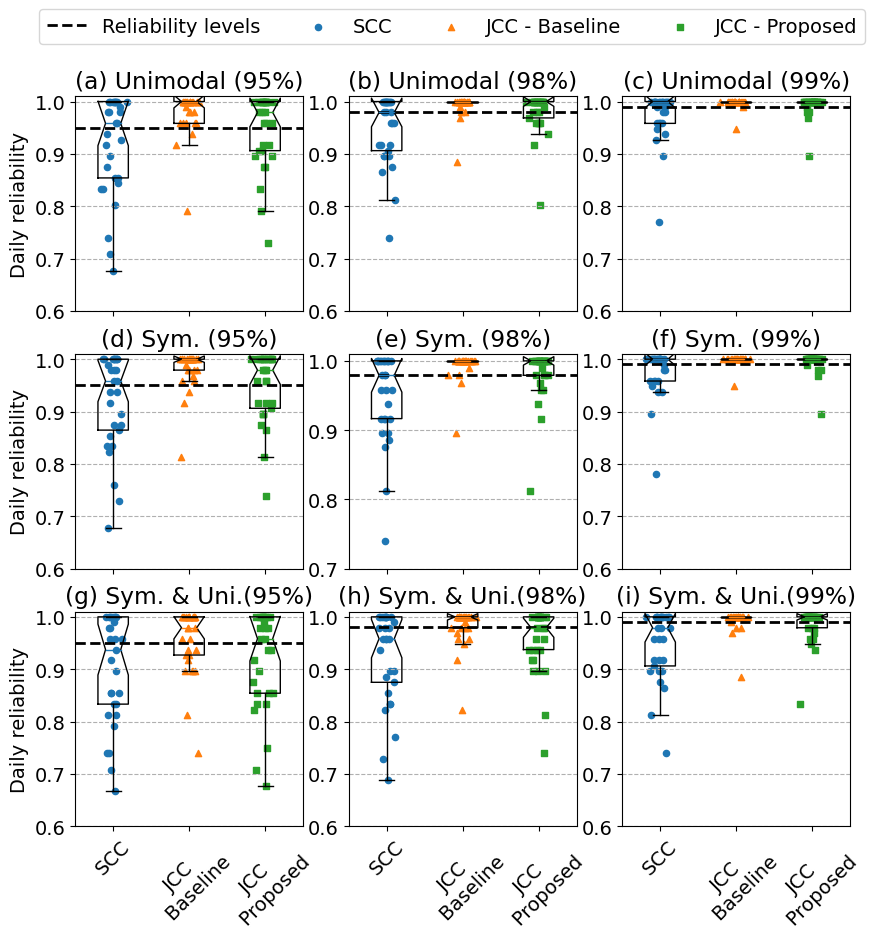}
        \caption{Summary of the full factorial robustness tests for three formulation methods. (Titles of each plot shows the experiment setting; Black dash lines show the reliability levels; Colored markers in the box plots are daily results in each experiment)}
    \label{robustness_test}
\end{figure}

We test three joint violation rates which are 0.05, 0.02 and 0.01 (i.e. the corresponding daily reliability requirements are 0.95, 0.98 and 0.99) and consider three aforementioned ambiguity sets, symmetric, unimodal and symmetric \& unimodal sets. For comparison, we include another two methods, SCC with all violation rates equal to the set reliability level, and the baseline case, JCC with the \textit{Bonferroni Approximation}. A full factorial experiment of these settings gives a total number of $3 \times 3 \times 3 = 27 $ robustness tests. \textcolor{black}{For each test, we use 10,800 forecast error samples (i.e., 30 samples for each 15-min time interval) from the historical sample pool.  Samples in each time interval are from different days and independent. Thus the power flow tests conducted in each time interval are considered to be independent.}

Fig. \ref{robustness_test} shows results from all robustness tests. Each subplot represents a combination of the daily reliability requirement and ambiguity set. Three box plots in each experiment from left to right show the results from SCC (blue circles), JCC using \textit{the Bonferroni Approximation} (orange triangles) and the proposed algorithm for JCC (green squares) respectively. Colored markers in each box plot are the daily results. Table III summarizes the average daily reliability in all robustness tests, and the reliability values closets to the set requirements are highlighted in \textbf{bold}.

\begin{table}[!h]
\caption{Summary of the average daily reliability (\%) in all robustness tests}
\label{average_reli}
\resizebox{0.49\textwidth}{!}
{
\centering
\begin{tabular}{@{}cccccccccc@{}}
\toprule
\textbf{Reliability}      & \multicolumn{3}{c}{\textbf{95\%}} & \multicolumn{3}{c}{\textbf{98\%}} & \multicolumn{3}{c}{\textbf{99\%}} \\ \midrule
\textbf{Cases}                                                            & (a)        &    (d)         & (g)  & (b)            &    (e)     &  (h)  & (c)          &   (f)       & (i) \\ \midrule
SCC                                                           & 91.5  &  91.9 & 89.9 & 94.6 & 94.9 & 92.3 &97.4 &97.5 & 94.6\\
JCC-B                                                               &97.8 & 98.1 &\textbf{95.3} & 99.3  & 99.4  & \textbf{98.2} & 99.8 & 99.8 & \textbf{99.3}\\ 
JCC-P                                                            &\textbf{94.3}   &\textbf{94.6}  &91.7 & \textbf{97.9}  &\textbf{98.1} & 95.4 &  \textbf{99.4} & \textbf{99.4} &98.4\\ \bottomrule
\end{tabular}}
    \begin{tablenotes}
      \footnotesize
            \item SCC: Single chance constraints; JCC-B: Joint chance constraints (Baseline); JCC-P: Joint chance constraints (Proposed)
    \end{tablenotes}
\end{table}

The SCC method fails to meet reliability requirements in all tests. More than half of its reliability outcomes are below the set requirements (i.e. black dash lines in Fig. 8) and the average daily reliability is also far lower than requirements (Table III). In contrast, JCC with the \textit{Bonferroni Approximation} gives an overly-conservative solution. In cases except (g) and (h), the majority of daily reliability outcomes are exactly or nearly 100\%. This makes its average reliability outcomes are highest and above the set requirements in all cases. The proposed algorithm can decide the optimal trade-off between the system reliability and operation cost. In each test, more than half of its daily reliability meets the requirements (i.e. the median of the box plot is all above the reliability threshold), and its average reliability values in six of nine cases are closest to the set reliability levels. In cases (g), (h) and (i), the underlying uncertainty distribution assumption is based on the unimodal \& symmetrical set, which is the smallest set among all three. The set can not fully incorporate all true distributions and impacts the reliability performance. The average values are below the reliability requirements. 

For three ambiguity sets, the rank based on the system reliability from the highest to the lowest is symmetric, unimodal and unimodal \& symmetric sets respectively. This result is aligned with the operation cost (Fig. \ref{summary}). When the set gets more constrained, the solution becomes less reliable and the operation cost becomes lower, and vice versa. In our case, the unimodal \& symmetric set is overly-constrained to describe the true distribution of solar forecast errors. However, even using the other two sets, there are a few discrete outliers in particular days which are costly to address, unless using an ambiguity set including all possible distributions. A feasible method for system operators could be to predict those particular days using the multi-year data and prepare the extra storage for those times.

\section{Conclusion}

The paper proposes a DR-JCC framework for microgrids considering solar generation uncertainty and utility contingencies. The framework models a networked microgrid with an islanding capability and smart load shedding during blackouts. Under imperfect solar forecasts, it optimizes chance constraints pertaining to the power flow, voltage control and battery limits \textit{jointly}, to decide the optimal trade-off between the operation cost and system reliability.

To find the optimized individual violation rates for the JCC problem, we propose a novel population-based evolutionary algorithm to optimize the decision variables and individual violation rates simultaneously. Results clearly show the proposed algorithm can effectively solve the problem non-conservatively. It can reduce the operation cost by around 50\% compared to the benchmark case, which has evenly individual violation rates (i.e. \textit{Bonferroni Approximation}). Moreover, the individual violation rates from the proposed method indicate the cost sensitivities of constraints. A higher violation rate means this constraint is more influential to the total cost. 

For solution robustness, we consider three ambiguity sets for solar power forecast error distributions, unimodal, symmetric and unimodal \& symmetric sets, based on empirical samples. We solve the model based on these set assumptions and then test the constraint violations with new forecast errors. Under the well-fitting ambiguity set assumption, the solution from the proposed method can control the system to closely meet the reliability requirements. The single chance-constrained formulation widely used in current research and practices, however, shows poor performance in securing reliability. 

\textcolor{black}{Future research will investigate the implementation of the convex AC OPF such as \cite{Gan2014} which incorporates active, reactive generation limits and power losses. This research will study how individual violation rates change with power flow conditions such as line congestion and heavy loading. }Another important area is to implement this approach in the transmission network operation and planning (e.g. \cite{Zhang2017}, \cite{Dvorkin2020a}) which can shift the paradigm in risk and reliability management especially under extreme weather and create greater value by saving million-scale reserve procurement costs.



\bibliographystyle{IEEEtran}
\bibliography{bibtex/IEEEconference}

\begin{IEEEbiography}[{\includegraphics[width=1in,height=1.25in,clip,keepaspectratio]{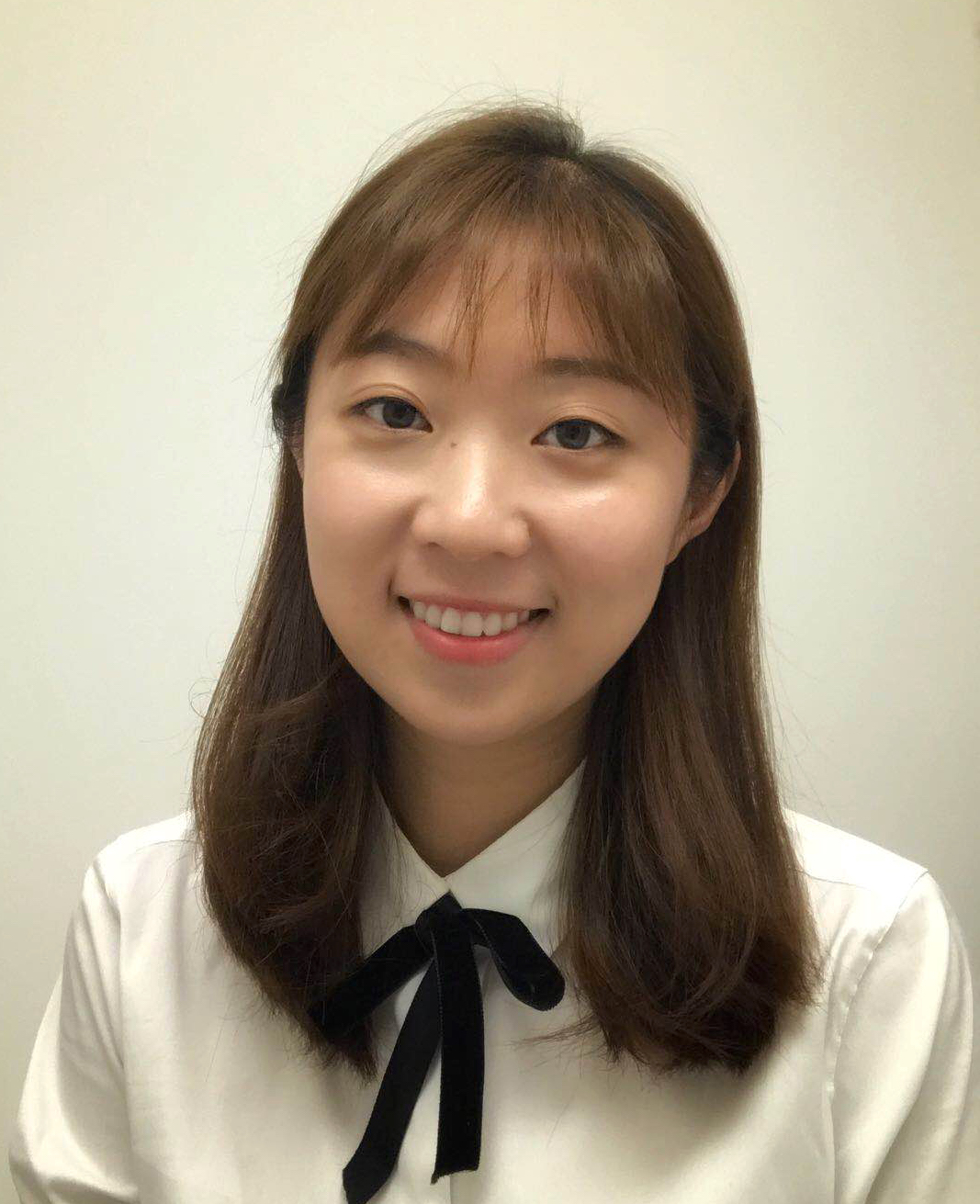}}]%
{Yifu Ding}
(Student Member, IEEE) received the B.Eng. degree in electrical and electronics engineering from the University of Edinburgh, UK in 2016 and the M.Sc. degree in sustainable energy futures from Imperial College London in 2017. She is currently pursuing a D.Phil. degree in engineering science with the Energy and Power Group, Department of Engineering Science, University of Oxford. Her research interests include machine learning applications, data-driven optimization and control for uncertainty management in power systems.
\end{IEEEbiography}

\begin{IEEEbiography}[{\includegraphics[width=1in,height=1.25in,clip,keepaspectratio]{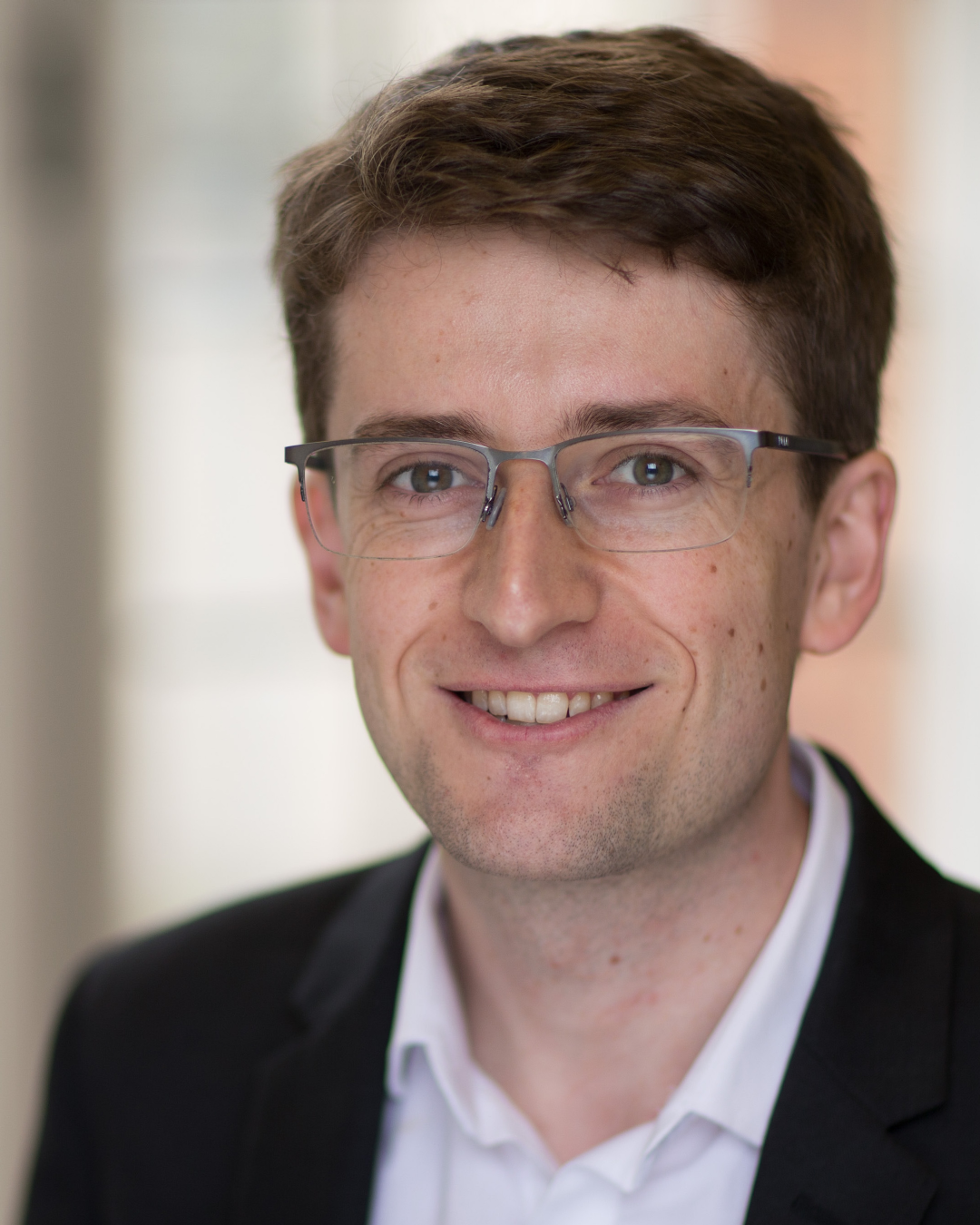}}]%
{Thomas Morstyn}
(Member, IEEE) received the BEng (Hon.) degree from the University of Melbourne in 2011, and the Ph.D. degree from the University of New South Wales in 2016, both in electrical engineering. He is a Lecturer of Power Electronics and Smart Grids with the School of Engineering, University of Edinburgh. His research interests include multi-agent control and market design for integrating distributed energy resources into power system operations.
\end{IEEEbiography}

\begin{IEEEbiography}[{\includegraphics[width=1in,height=1.25in,clip,keepaspectratio]{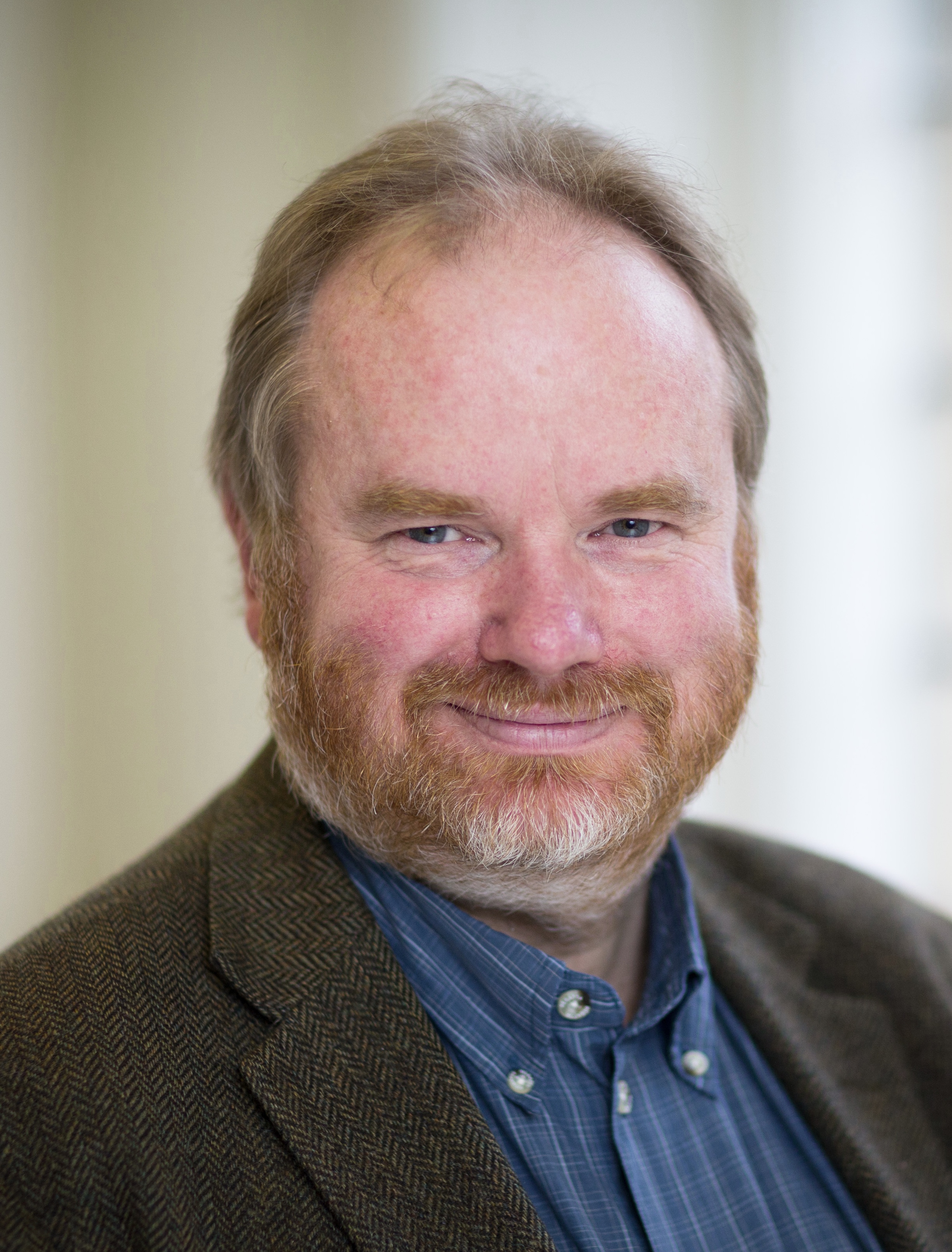}}]%
{Malcolm McCulloch}
(Senior Member, IEEE)
received the B.Sc. (Eng.) and Ph.D. degrees in
electrical engineering from the University of the
Witwatersrand, Johannesburg, South Africa, in 1986
and 1990, respectively. In 1993, he joined the
University of Oxford, Oxford, U.K., to head up
the Energy and Power Group, where he is currently an Associate Professor with the Department
of Engineering Science. His work addresses transforming existing power
networks, designing new power networks for the developing world, developing new technology for electric vehicles, and developing approaches to
integrated mobility.
\end{IEEEbiography}

\end{document}